\documentclass[sigconf]{acmart}

\usepackage{graphicx}
\usepackage{listings}
\usepackage{multirow}
\usepackage{varwidth}
\usepackage{xcolor}
\usepackage{amsthm}
\usepackage{amsmath}
\usepackage{float}
\usepackage{enumitem}
\usepackage{subcaption}

\AtBeginDocument{%
  \providecommand\BibTeX{{%
    \normalfont B\kern-0.5em{\scshape i\kern-0.25em b}\kern-0.8em\TeX}}}

\copyrightyear{2020}
\acmYear{2020}
\setcopyright{acmcopyright}
\acmConference[MTD '20] {7th ACM Workshop on Moving Target Defense}{November 9, 2020}{Virtual Event, USA}
\acmBooktitle{7th ACM Workshop on Moving Target Defense (MTD '20), November 9, 2020, Virtual Event, USA}
\acmPrice{15.00}
\acmDOI{10.1145/3411496.3421228}
\acmISBN{978-1-4503-8085-0/20/11}

\begin{document}
\fancyhead{}

\title{Range and Topology Mutation Based   Wireless Agility}

\author{Qi Duan}
\affiliation{%
  \institution{Carnegie Mellon University}
  \country{U.S.A}
}
\author{Ehab Al-Shaer}
\affiliation{%
  \institution{Carnegie Mellon University}
  \country{U.S.A}
}
\author{Jiang Xie}
\affiliation{%
  \institution{University of North Carolina at Charlotte}
  \country{U.S.A}
}


\begin{abstract}
Wireless is a key component in most of today's network infrastructures. Yet, it is highly susceptible to network attacks because wireless communication and infrastructure, such as Access Point (AP) and clients, can be easily discovered and targeted. Particularly, the static nature of the wireless AP topology and its configuration offers a significant advantage to adversaries to identify network targets and plan devastating attacks such as denial of service or eavesdropping. This is critically important in hostile military environment in which soldiers depend on wireless infrastructure for communication and coordination.

In this paper, we present formal foundations for two  wireless agility techniques: (1) Random Range Mutation (RNM) that allows for periodic changes of AP coverage range randomly, and (2) Random Topology Mutation (RTM) that allows for random motion and placement of APs in the wireless infrastructure. The goal of these techniques is to proactively defend against  targeted attacks (e.g., DoS and eavesdropping) by forcing the wireless clients to change their AP association randomly. We apply
  Satisfiability Modulo Theories (SMT) and Answer Set Programming (ASP) based constraint
   solving methods that allow for optimizing wireless AP mutation while maintaining  service requirements including coverage, security and energy properties under incomplete information about the adversary strategies. 
Our evaluation validates the feasibility, scalability, and
effectiveness of the formal methods based technical approaches.
\end{abstract}



\copyrightyear{2020} 
\acmYear{2020} 
\acmConference[MTD'20]{7th ACM Workshop on Moving Target Defense}
\acmBooktitle{7th ACM Workshop on Moving Target Defense (MTD'20)}
\acmPrice{15.00}
\acmDOI{10.1145/3338468.3356830}
\acmISBN{978-1-4503-6828-5/19/11}

\maketitle

\section{Introduction and Motivation}
Wireless communications has become a key component in  most of today's network infrastructures, both in military and commercial applications. Indeed, the surge in the demand for wireless communication has led to a major revolution in the wireless landscape. In the next few years, it is anticipated that a viral deployment of small-sized wireless base stations, known as \emph{small cell base stations} (SBSs) will occur at a massive scale in order to lay the foundations of the next-generation  5G networks which can sustain the growth in the demand for the wireless capacity~\cite{IG00,JA02,JA03}. SBSs include a variety of access points (APs) such as picocells, femtocells, microcells, and macrocells, varying in coverage, capacity, and capabilities~\cite{JA03}. One of the key features of SBSs is their low-cost, low-coverage nature and their ability to be connected to any existing backhaul wired network such as DSL~\cite{IG00}. Such SBSs must co-exist with other wireless infrastructures such as WiFi APs, which are overlaid on top of this cellular architecture.
This massive network densification  will lead to a major paradigm shift from current controlled, homogeneous wireless systems composed of sophisticated base stations or APs to decentralized, large-scale heterogeneous networks composed of cheap and small access points.
A typical wireless network involves attaching several APs to a wired network and then providing wireless access to the LAN/WAN or another wireless network. The wireless access points such as SBSs are managed by a  controller that handles automatic adjustments to radio frequency  power, channels, authentication, and security. The provision of security services in the wireless context faces a set of challenges that are specific to this new technology,
which includes: a) small infrastructure nodes such as SBSs or small APs are more vulnerable to hacking than expensive
 and large base station towers, b) wireless communication and infrastructures (such as SBSs, WiFi access points, and even wireless clients) can be easily discovered and targeted, and c) SBSs and similar APs must connect to third party wired backhauls such as DSL, which can increase their vulnerability. More importantly, the static nature of the wireless AP topology in all existing infrastructure (cellular, unlicensed WiFi, or ad hoc) and its configuration offers a significant advantage to adversaries to identify network targets and plan devastating attacks such as denial of service or eavesdropping. This is critically important in hostile military environments in which soldiers depend on wireless infrastructure for communication and coordination.
 There are  a number of existing dynamic protocols in wireless networks, such as randomized multi-path routing and wireless device status scheduling. The main objective of these protocols is to increase network resiliency in case of dynamic ad hoc network topology~\cite{LG01,YKT03,PH02,MKD07}. However, in these protocols, the wireless device status and route selection are predictable, which makes the network vulnerable to the adversary. The multipath algorithm in~\cite{SKL10} can generate randomized multipath routes that are also highly dispersive and energy efficient in wireless sensor networks, but it only targets  single black hole attacks.

In this paper, we explore   wireless network mutation techniques that
consider the adversary actions in order to maximize the benefit of network agility. Note that
a complete random mutation could be too costly and even disruptive for network services. In addition, valid mutation is constrained by the availability of network resources such as signal power, number of APs, distribution
 of wireless clients, etc. Thus, one of the main challenges in this research is to develop a formal framework for effective wireless agility (e.g., low system and client overhead), while maintaining coverage, energy, and security (e.g., access control) constraints.
We investigate formal foundations for two AP wireless agility techniques: (1) Random Range Mutation (RNM) that allows for periodic changes of AP coverage range randomly, and (2) Random Topology Mutations (RTM) that allows for random motion and placement of APs in the wireless infrastructure. The goal of these techniques is to proactively defend against reconnaissance and targeted attacks (e.g., DoS and eavesdropping attacks) by forcing the wireless clients to change their AP association and forwarding routes randomly . 
We apply an SMT and ASP
based approach to solve the wireless agility planning problems.
SMT is a powerful tool to solve constraint satisfaction problems arise
in many diverse areas, including software and hardware verification, type
inference, extended static checking, test-case generation, scheduling, planning, graph
problems, etc.~\cite{BM09}.
An SMT instance is a formula in first-order logic with equalities involving uninterpreted functions (EUF)~\cite{DLL62,DP60,Ganesh07}.
SMT solvers can determine the values of the Boolean 
and arithmetic   variables that make a set of 
logic and arithmetic formula  satisfying.
ASP\cite{asp1, asp2} is a form of declarative programming oriented towards difficult search problems. ASP is ideal for solving logic based
constraint satisfaction problems, and it is more expressive than Prolog
and Mincost SAT, and 
its language is more simple and elegant than SMT, and sometimes more efficient.
For example, for many one-player games 
 one can use ASP to provide the definition and constraint of the game
 and find the satisfiable solutions, without providing the actual 
 search algorithm. 

 For example, the SMT instance with the following two constraints:
\[ (x+y<2) \lor (x-2y>0) \]
\[  x \leq 1\]
  is satisfying, because it can be satisfied with the assignment $x=0,y=0$.

Comparing with SAT, SMT provides a much richer modeling language using EUF~\cite{DP60,GJ90}.
Yet modern SMT solvers can solve formulas with hundreds of thousands variables
and millions of clauses~\cite{MB09}.

To the best of our knowledge, we are the first to propose AP
mutation based wireless agility techniques.  We believe that our work will enable the implementation of mutable wireless networks and motivate other kinds of moving target defense in wireless networks.

The rest of the paper is organized as follows. Section~\ref{sec:model}
discusses the network and threat model used in the paper.
Section~\ref{sec:prob} and ~\ref{sec:tech}
 present the problem definitions and technical approaches
 respectively. Implementation details are discussed
 in Section~\ref{sec:imp}. Evaluation results are presented in
 Section~\ref{sec:eval}. Section~\ref{sec:related} presents
 related works and Section~\ref{sec:con} concludes the paper.

\section{System and Adversary Model}
\label{sec:model}
The network and threat model in this paper conform
with the following conditions:

(1) The wireless network has a number of APs, which may
have dynamic configurations. An individual AP can provide connection to
 devices in its wireless radio range and is connected to one or more
gateways or other APs (interconnected APs are similar
as wireless ad hoc networks). A wireless device may be inside the
range of multiple APs.
 The AP mutation (through configuration change) is controlled and regulated
by a central controller.

(2) Every AP can obtain its location in movement through
GPS, which is the common situation for military wireless
networks.

(3) An adversary has limited amount of
resources that are deployed in the network.
We consider two capabilities for the adversary: eavesdropping
and jamming. An adversary can deploy a number of
attacking nodes in the network.  Since APs in
 overlapped ranges will use different radio frequencies, we assume that every attacking node
has the ability to eavesdrop or block (through jamming)
 only one AP in its range in a given time interval.
However,
she does not control certain elements such as mobility of the
 APs or modification/addition of the
hardware of the captured APs. This assumption
is perfectly legitimate since our model considers
that the adversary does not know all the details of
the network and it will exponentially increase the
cost of gathering these details.

\section{Problem Definition}\label{sec:prob}

\noindent{\bf Random Range Mutation (RNM).} Randomly changing AP range forces wireless clients to switch their associated APs and routes. Thus, eavesdropping or DoS attackers who are targeting specific APs or clients will be confused since APs (as well as client associations) appear and disappear randomly and frequently.  Also adversaries can not make any assumption about the client association with APs based on physical location. For instance, a client could be associated to an AP that is further than another AP based on the assigned range, which also dynamically changes.
We assume that there is enough APs that can collaboratively cover the target area but in different ways.
Note that one special case of AP range mutation is that multiple APs that are
in close physical locations turn on/off alternately. Turning an AP off is
equivalent to reducing its range to 0.

The RNM problem can be formally defined as follows. Suppose that the set of users in the wireless network is denoted as   $\mathcal{P}=\{p_1,\ldots,p_z\}$.
In this network, a set $\mathcal{N}$ of  $N$ range adjustable APs  $\{s_1,s_2,\ldots, s_N \}$ is present. Every AP $s_i \in \mathcal{N}$ ($1 \leq i \leq N$) has $g_i$ possible ranges, denoted by the set  $\mathcal{S}_i=\{\alpha_{i1}, \alpha_{i2},\ldots ,\alpha_{ig_i}\}$. When $s_i$ is set to be in range  $\alpha_{ij}$ ( $1 \leq j \leq g_i$), it covers
a subset  $\mathcal{Y}_{ij}$ of users within the set $\mathcal{P}$, and the energy consumption rate is given by a function $f_i(\alpha_{ij})$. For an effective RNM, our goal is to find a range mutation schedule (range value, starting and ending time) for each AP that satisfies the following constraints:

\begin{itemize}
\item{\em Coverage constraint:} coverage of wireless users must be maintained.
\item{\em Unpredictability constraint:} increasing unpredictability by minimizing the similarity between new and old  mutation.
\item{\em Capacity constraint:} the number of users that handled by every AP should not exceed the AP's capacity.
\item{\em Energy constraint:} optimize energy consumption, because different range for the wireless APs will lead to different energy consumption rate.
\end{itemize}

\noindent{\bf Random Topology Mutation (RTM).} The topology mutation is based on mobile APs that change their positions in random manner but with constraints. When a wireless AP changes its position, some wireless users may need to change its associated AP, and change  forwarding routes accordingly.  By moving the positions of the APs, we can achieve better security with limited number of APs. The objectives of random topology mutation are (1) to make wireless infrastructure resilient against infrastructure  reconnaissance attacks, (2) to defend against physical attacks and increase
 the resilience to wireless device failures, which is particularly
 important for military applications, (3) to make eavesdropping attacks that are targeting
   specific clients extremely difficult as the attacker has to chase not
  only the mobility of client but also the mobility of the AP, and (4) to
  make wireless resilient against internal attacks in case some APs are
  compromised, because (a) one AP alone cannot focus on tracking or
  attacking (e.g., eavesdropping) a specific client as each AP will be
  constantly moving and the harm will be naturally distributed, and (b) the agile AP enables clients to compare the performance across various APs and potential detect malicious APs (e.g., one AP drops packets maliciously).

The topology mutation can be done in two scheduling
phases. The first phase is to determine the satisfiable deployment of the APs for the next interval,
and the second phase is to determine the
detailed moving steps to move the APs from the current deployment to the new deployment.
For the first phase, the following constraints should be satisfied for random AP deployment:

\begin{itemize}
\item{\em Coverage constraint:} coverage of wireless users must be maintained.
\item{\em Unpredictability constraint:} increasing unpredictability by minimizing the topology similarity between new and old topology.
    \item{\em Capacity constraint:} the number of users that handled by every AP should not exceed the AP's capacity.
     \item{\em Connectivity constraint:} the AP network should be
     a connected graph.
\end{itemize}

For the second phase, the major constraints are
\begin{itemize}
\item{\em Step constraint:} the total number of
moving steps needed for the transition should be no more than some threshold value.
\item{\em Energy constraint:} the total energy cost for  moving from the old deployment to the new deployment should be less than some threshold.
    \item{\em Connectivity constraint:} during every step 
    of the movement, the AP network should be
     a connected graph.
\end{itemize}

\section{Technical Approaches}
\label{sec:tech}
\subsection{Preliminary Formalization and Constraints}\label{sec:cons}
\subsubsection{Random Range Mutation}
\label{subsec:RNM}
Random range   mutation  can be pre-calculated and staged in wireless APs  in advance and activated based on moving target requirements. We apply
   SMT  to solve random range   mutation.

The following equation  specifies the energy consumption upper bound ($E_i$ ) for every AP $s_i$ ({\em energy constraint}):
\begin{equation}
\label{equ:energy}
 \displaystyle\sum_{1\leq j \leq T}\omega_{ij}f_i(\omega_{ij}) \leq E_i,  \,\,  \forall i \in [1..N]
\end{equation}

Here the variable $\omega_{ij}$ denotes the range of AP $s_i$ at time interval $j$.
The  range can be one of $\alpha_{i1},\ldots, \alpha_{ig_i}$, which can be represented by
natural numbers. For example, we can use numbers $1,2,\ldots,g_i$ to represent
  $\alpha_{i1}, \alpha_{i2}, \ldots, \alpha_{ig_i}$, respectively.

To set the choice of values of variable $\omega_{ij}$, we need the following equation.
\begin{equation}
\label{equ:choice}
 \omega_{ij} \in \{\alpha_{i1}, \alpha_{i2},\ldots ,\alpha_{ig_i}\}, \,\, \forall i \in [1..N], \forall j \in [1..T]
\end{equation}

The following equation guarantees that every user should be covered by at least one
AP in any time interval ({\em coverage constraint}).

\begin{equation}
\label{equ:cover} (\displaystyle\bigvee_{(i,u)\in\{ (i,u)|p_k \in \mathcal{Y}_{iu} \} }
              \omega_{ij} = \alpha_{iu} ),   \,\,  \forall k \in [1..z], \forall j \in [1..T]
\end{equation}

The next equation guarantees that the range  of every AP in any time interval will be different from
the previous time interval ({\em unpredictability constraint}).

\begin{equation}
\label{equ:unpredict}
 \omega_{ij} \neq \omega_{i(j+1)} , \,\, \forall i \in [1..N], \forall j \in [1..T-1]
\end{equation}

Note that the unpredictability constraint can be flexible. One can require that
the range of an AP must change over a number of time intervals (not necessarily  two consecutive
intervals).

 Suppose the  capacity of AP $s_i$ (the maximum number of wireless users that can connect to it) is $Q_i$. Then
  the {\em capacity constraint}
can be formalized as follows 
\begin{equation}
\label{equ:assign}
 \displaystyle \sum_{1\leq i \leq N}v_{ijk} = 1, \,\, \forall j \in [1..T], \forall k \in [1..z]
\end{equation}

\vspace{-.2in}
\begin{eqnarray}
\label{equ:in_range}
  (v_{ijk} = 1)  \Rightarrow \exists u (  (p_k \in \mathcal{Y}_{iu} )
    \land (\omega_{ij} = \alpha_{iu}) ), \nonumber \\
   \forall i \in [1..N], \forall j \in [1..T], \forall k \in [1..z]
\end{eqnarray}

\vspace{-.2in}
\begin{equation}
\label{equ:capacity}
 \displaystyle \sum_{1\leq k \leq z}v_{ijk} \leq Q_i, \,\, \forall i \in [1..N],\forall j \in [1..T]
\end{equation}


In the above equation, $v_{ijk}=1$ means that user $p_k$ is assigned to AP $s_i$
in interval $j$. Equation~\ref{equ:assign} guarantees that every user will be assigned with one AP. Equation~\ref{equ:in_range} guarantees
  that when a user node is assigned by an AP, it should
  be inside the active range of the AP in the current interval.
   Equation~\ref{equ:capacity} guarantees that every AP will not exceed its capacity.

\subsubsection{Random Topology Mutation}
\label{subsec:RTM}
We apply ASP to solve random topology mutation since it is more
convenient to define connectivity or reachability related 
constraints. ASP is ideal to solve this kind of constrained 
movement problems since it is very convenient to define
movement related constraints in ASP and the program of ASP
is  simpler than SMT.

In RTM, the following formalization
 guarantees that every user should be covered by at least one AP for the deployment in the next interval ({\em coverage constraint}).

\begin{equation}
\label{equ:cover_topology}
 (\displaystyle\bigvee_{(i,j)\in\{ (i,j)|p_k \in \mathcal{Y}_{ij} \} }
              \omega_i = \alpha_{ij} ),   \,\,  \forall k \in [1..z]
\end{equation}

In the above equation,  the variable $\omega_i$ denotes the position of AP $s_i$
in the next deployment. The  choice of values of variable $\omega_i$ is defined by
\begin{equation}
\label{equ:choice_topology}
 \omega_i \in \{\alpha_{i1},\ldots ,\alpha_{im}\}, \,\, \forall i \in [1..N]
\end{equation}

In the above equation,  $\{\alpha_{i1}, \ldots ,\alpha_{im}\}$ are the possible locations that
AP $s_i$ can move to.

The {\em unpredictability constraint} can be formalized as follows:
 \begin{eqnarray}
\label{equ:capacity_topology}
  \displaystyle\sum_{1 \leq i \leq n} \Delta_i \geq \delta \\
 (\Delta_i = 1 ) \Leftrightarrow  (\omega_i = \eta_i), \,\, \forall i \in [1..N] \\
  \Delta_i \in \{0, 1\}  , \,\, \forall i \in [1..N]
\end{eqnarray}

In the above equations,  $\delta$ is the overlap threshold (the minimum number of
APs that should change location), and $\eta_i$ is the old position of AP $s_i$.
The unpredictability constraint guarantees that the number of
moved APs in consecutive intervals should exceed a certain threshold.

The {\em Connectivity constraint} can be formalized as follows
\begin{align}
  Reachable(s_1, s_i), \,\, \forall i \in [2..N]\\
  ( Reachable(s_1, s_j) \land neighbor(s_i, s_j) )
     \rightarrow  Reachable(s_1, s_i)
\end{align}

Note that here we define connectivity as that node 1 can 
reach every other node. Reachability is defined recursively based on that a
node can reach any of its neighboring nodes.

The capacity constraint can be defined similarly as that in random range mutation.

In the second phase of topology mutation scheduling, suppose the maximum number of
steps (one step means
moving from a location to an adjacent location)
 required for the transition is $b$,  we can formalize the {\em step constraint} as follows:
   \begin{eqnarray}
\label{equ:time_topology}
(\omega_{i1} = \eta_i ) \land (\omega_{ib} = \overline{\eta_i} ), \forall i \in [1..N]\\
  (\omega_{i(j+1)} = \omega_{ij}) \lor ( \omega_{i(j+1)} \text{ neighbor to } \omega_{ij} ), \nonumber \\
      \forall i \in [1..N] , \forall j \in [1..b-1]
\end{eqnarray}

In the above equations,  $\eta_i$ is the position of $s_i$ in the original deployment, and $\overline{\eta_i}$
is the  position of $s_i$ in the new deployment. $\omega_{ij}$
denotes the position of $s_i$ in step $j$. From step 1 to step $b-1$, we
require that every movable AP should either remain in the previous position
or move to a neighboring position.

Assume that every move in one step consumes one unit energy, we can formalize the {\em energy constraint} as follows:
\begin{eqnarray}
\label{equ:energy_topology}
  \displaystyle \sum_{1 \leq j \leq b-1} \zeta_{ij} \leq E_i, \forall i \in [1..N]\\
  (\omega_{i(j+1)} \neq \omega_{ij}) \Leftrightarrow (\zeta_{ij}=1), \,\,  \forall i \in [1..N], \forall j \in [1..b-1] \\
  \zeta_{ij} \in \{0,1\}  , \,\,  \forall i \in [1..N], \forall j \in [1..b-1]
\end{eqnarray}

In the above equations,  $E_i$ is the threshold for energy consumption
for AP  $s_i$ and $\zeta_{ij}$ denotes whether AP $s_i$ moves in step $j$ or not.

Connectivity constraints can be defined similarly as in the first phase.

\section{Implementation}
\label{sec:imp}

\subsection{Protocol Overview}

We propose a protocol to implement the proposed AP mutation schemes. In our proposed network architecture, a centralized controller is connected to all the APs in the network. This centralized controller acts as a radio resource coordinator across the network and takes care of issues related to AP coverage range mutation and random movement and placement. Note that the centralized controller is not necessarily  an additional separate device. Instead, the functions of the controller can be implemented in an existing AP, and this AP becomes the controller of all APs.

The centralized controller gathers the measured signal and resource utilization statistics from all the APs via the backbone network connecting all the APs using Simple Network Management Protocol (SNMP) \cite{SNMP}. SNMP also provides security related functions such as user authentication and message encryption \cite{SNMPv3}. Most enterprise-class APs can support SNMP \cite{Matsunaga04}. APs collect signal characteristics from wireless users. Based on IEEE 802.11k \cite{80211k} radio resource measurement, signal characteristics can be obtained directly from the wireless network.

The controller and APs periodically collect the required information for AP mutation. The controller calculates the optimal range mutation schedule or the optimal AP deployment and movement schedule and sends decisions to APs.  Due to the dynamics in the RF environment, signal characteristics, traffic load, and interference intensity are time-variant. As a result, the decision-making is updated periodically in order to reflect the influences of the time-varying environment. The signaling overhead caused by the proposed AP mutation schemes comes from the periodic measurement collection of the AP load and signal statistics from all APs to the controller. However, the traffic load at APs does not vary frequently, if users are not highly mobile.
In addition, it is important to remark that the periodic data-collection does not imply measuring instantaneous small-scale multipath signal characteristics which are very time-sensitive. Instead, measurements should be targeted at capturing large-scale changes in signal characteristics due to variations in traffic pattern, user mobility, interference sources, and interference mobility.
Thus, the signaling overhead that results from the periodic updating can be kept at a reasonable level. Generally speaking, the shorter the measurement interval, the more often the optimization decision is updated, the better performance the system can achieve in terms of wireless agility and load balancing based on the most recent traffic load and interference information, but the higher the signaling overhead.

The proposed protocol includes three steps. First, the controller collects  the AP coverage information as well as the overall traffic load distribution and user distribution at all APs. Second, the controller finds the optimal mutation schedule (either range mutation or topology mutation) for each AP. In other words, the controller decides which AP should use which transmission power level to cover which range or which AP should move to which location. This optimal mutation schedule satisfies all the constraints. Finally, the controller sends control decisions to APs to instruct them on how to update their coverage ranges or locations. Therefore, by the control of the centralized controller on restricting the range or topology of APs, the network agility can be maximized, while user coverage, network capacity, and energy consumption
 constraints are not compromised.

\subsection{Handoff in AP Mutation}

One of the major challenges in this work is to maintain existing connections during the mutation. To achieve this, we must switch the association of the wireless node from one AP to another AP. To achieve a smooth switch, one can  use soft handoff. Note that the soft handoff we apply here is a kind of proactive handoff caused by mutation. In the traditional cellular network,  handoff is caused by the movement of the phone users, or  when the capacity for connecting new calls of a given cell is used up, or user behavior changes. For example, when a fast-traveling user connected to a large umbrella-type of cell stops,  the call may be transferred to a smaller  cell in order to free capacity on the umbrella cell for other fast-traveling users and to reduce the potential interference. For the proactive handoff caused by range mutation, the handoff happens even when the wireless user is not moving.

We can use the IAPP (Inter-Access Point Protocol)~\cite{iapp} for soft handoff. IAPP (standardized in IEEE Std 802.11f) is the IEEE standard for inter-access point communication.  IAPP  enables (fast) link layer re-association at a new access point in the same hotspot.


\section{Related Works}
\label{sec:related}
Cyber agility is an important research topic
in recent years. It
provides a proactive defense to deter many attacks including worm, botnet, DoS and renaissance attacks with the presence of uncertainties of the timing and types of attacks. The work in~\cite{ADJ12} presents the framework
of random host IP mutation (RHM). In RHM, moving
target hosts are assigned virtual IP addresses that change
randomly and synchronously in a distributed fashion over time
without disrupting active connections.
IPv6 based mutation is investigated
in~\cite{ipv611} to leverage the
immense address space of IPv6.   The work in~\cite{CNS13}
presents Random Route Mutation (RRM) that
enables changing randomly the route of
the multiple flows in a network simultaneously to defend against
reconnaissance, eavesdrop and DoS attacks, while preserving
the required operational and security properties.

 Agility in wireless/mobile networks is particularly interesting
 due to the intrinsic dynamic nature  and  special vulnerabilities
  of the wireless/mobile
 networks. The Software-defined radio (SDR)~\cite{asp1} is essentially
 an agile radio communication system which is flexible  to avoid the limited spectrum  of previous kinds of radios. The work in~\cite{GMMT13}
  applies a moving target defense to all nodes within a mobile-enabled system
  to provide  security for critical nodes in the  network.
The work in~\cite{chin16} presents a Moving Target Defense  approach to concealing the location of a base station within a wireless 
sensor network (WSN). The approach 
uses multiple base stations to serve a WSN where one of the multiple base stations is chosen to serve the WSN in a fixed duration of time. This selection
changes over a time to provide obscurity of the base station location information. 
MPBSD is an effective MTD approach to secure base stations for a WSN in term of sensory performance such as end-to-end delay.
The work in~\cite{ge20} proposes an integrated defense technique to achieve intrusion prevention by leveraging cyber deception and moving target defense
together. The effectiveness and efficiency of the proposed technique is shown analytically based on a graphical security model in a software defined networking (SDN)-based IoT network. 

  Approaches for load balancing in a wireless local area network
  can be classified into two categories. One is association control
  through which the network re-distributes client associations
  among APs more or less uniformly so that no one AP is
  unduly overloaded~\cite{HF04,MZPPH08,BJL07}. The other is capacity
  control through which the network adjusts the maximum
  allowable throughput of each AP so that heavy-loaded APs can have
  more capacity to support users~\cite{Matsunaga04}. However, these
  works are not from the security perspective.

\section{Conclusion and Future Work}
\label{sec:con}
In this paper, we investigate
   two  wireless agility techniques: (1) Random Range Mutation (RNM) that allows for periodic changes of AP coverage range randomly, and (2) Random Topology Mutations (RTM) that allows for random motion and placement of APs in the wireless infrastructure.
   We apply
  SMT  based formal framework
   to schedule the satisfying wireless AP mutation
     considering coverage, security, and energy
     constraints. Our evaluation shows that the
     AP mutation techniques can effectively defend against
     eavesdrop and DoS attacks, with reasonably  performance degradation.
     The SMT based formalization can solve
     mutation scheduling in topologies as large as
     2500 vertices, and the   throughput reduction is less than
     2\%. Compared with the case of no mutation, the percentage
     of compromised flows can decrease by more than 90\%.

The AP  mutation should also maintain the connectivity among the APs
if they form an ad hoc network. In the current formalization
we do not include connectivity constraint
since we assume there are a large number of APs
and  connectivity can be easily satisfied even
if a percentage of APs are inactive in a given interval. If there
is not enough number of APs or the percentage
of inactive  APs is large in a given interval, then
we need to consider connectivity constraints.  This
is  one direction of future work.

     Another important limitation of the
     presented approaches is that the action of APs may directly impact the
      the adversary strategies. For example, when an AP changes its coverage range or location,
       an adaptive adversary may also react accordingly by changing its own locations
        or attacking targets.
     In the future, we plan to
     extend the research to game theory based wireless agility
     to consider the mutual interaction between defenders
     and attackers.

\bibliographystyle{plain}
\bibliography{main}

\end{document}